%% file: ms.tex
\documentclass[graybox]{svmult}

\usepackage{mathptmx}       % selects Times Roman as basic font
\usepackage{helvet}         % selects Helvetica as sans-serif font
\usepackage{courier}        % selects Courier as typewriter font
\usepackage{type1cm}        % activate if the above 3 fonts are
\usepackage{makeidx}         % allows index generation
\usepackage{graphicx}        % standard LaTeX graphics tool
\usepackage{multicol}        % used for the two-column index
\usepackage[bottom]{footmisc}% places footnotes at page bottom

\makeindex             % used for the subject index
\def\uu{4U\,0142+614}
\def\ee{1E\,1048-5937}
\def\kes{1E\,1841-045}

\def\rxs{1RXS\,J1708-4009}
\def\ea{1E\,2259+586}
\def\xte{XTE\,J1810-197}

\begin{document}

\title*{The magnetar emission in the IR band: the role
of magnetospheric currents}
\author{Silvia Zane, Luciano Nobili and Roberto Turolla}
\authorrunning{The magnetar emission in the IR band}
 \institute{Silvia Zane \at Mullard Space Science Laboratory,
University College of London, Holmbury St Mary, Dorking, Surrey,
RH5 6NT, UK, \email{sz@mssl.ucl.ac.uk} \and Luciano Nobili,
Roberto Turolla \at Dept. of Physics, University of Padova, Via
Marzolo 8, 35131 Padova, Italy, \email{luciano.nobili@pd.infn.it,
roberto.turolla@pd.infn.it}}
%
% Use the package "url.sty" to avoid
% problems with special characters
% used in your e-mail or web address
%
\maketitle

\abstract*{qui abstract}

\abstract{ There is a general consensus about the fact that the
magnetar scenario provides a convincing explanation for several of
the observed properties of the Anomalous X-ray Pulsars and the
Soft Gamma Repeaters. However, the origin of the emission observed
at low energies is still an open issue. We present a quantitative
model for the emission in the optical/infrared band produced by
curvature radiation from magnetospheric charges, and compare
results with current magnetars observations.}

\section{Introduction}
\label{sec:1} The Anomalous X-ray Pulsars and the Soft Gamma
Repeaters (AXPs and SGRs) are peculiar X-ray pulsators which
exhibit erratic phases of bursting/outbursting activity (e.g.
Mereghetti  2008). There is now a wide consensus that these
sources indeed host an ultra-magnetized NS, or magnetar (Duncan \&
Thompson 1992),  with a magnetic field $B\approx
10^{14}$--$10^{15}\ {\rm G}$, well in excess of the quantum
critical field $B_{Q}\simeq 4.4\times 10^{13}\ {\rm G}$.

The magnetar scenario provides a quite coherent interpretation of
the observational properties of SGRs/AXPs, in particular of both
their bursting and persistent emission at X/$\gamma$-ray energies
(Wood \& Thompson 2006). In particular, it has been suggested that
the magnetosphere of a magnetar is twisted ($\nabla\times{\mathbf
B}\neq 0$), in which case the density of magnetospheric currents
is substantial and the observed soft X-ray spectrum ($<10$~keV)
can be explained in terms of resonant cyclotron scattering (RCS)
of thermal surface photons onto mildly relativistic particles
flowing along the closed field lines (see Zane et al. 2009 for a recent
application, and references therein).

Observations at IR/optical wavelengths led to the discovery of
faint ($\rm K\sim 19$--21), and in some cases variable IR
counterparts to several AXPs/SGRs (see again Mereghetti 
2008). In two sources optical pulsations at the X-ray period were
found (Kern \& Martin 2002, Dhillon et al. 2009). However, the
origin of the IR/optical emission is not clear as yet. The
proposed scenarios involve emission/reprocessing by a fossil disk
(see e.g. Perna et al. 2000), or non-thermal emission from the
inner magnetosphere (Eichler et al. 2002, Beloborodov \& Thompson
2007, hereafter BT07). Although this latter option appears
promising, no detailed computations have been presented as yet. In
this paper we reassess this issue, and provide a quantitative
estimate of the IR emission expected from curvature radiation.

\begin{figure}[b]
\sidecaption[t]
\includegraphics[scale=.40]{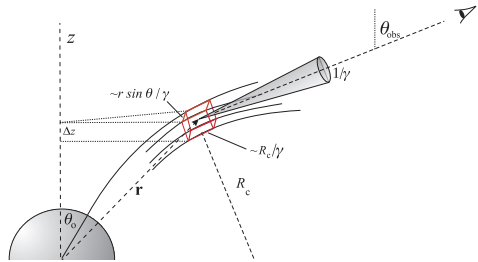}
\caption{ Geometry of the curvature emission. See text for
details.} \label{volumi}
\end{figure}

\section{Pair production in the inner magnetosphere}
\label{sec:2}

In the inner magnetosphere, the magnetic field is so strong that
the conversion of high-energy photons produced by RCS into pairs
may give rise to a quasi-equilibrium configuration in which the
$e^\pm$ Lorentz factor is locked at about the threshold value
required to ignite the pair cascade ($50 < \gamma < 1000$, BT07).
In order to estimate the spatial extension of this region, we
proceed as follows. We assume that primary photons have a typical
energy $\epsilon\sim 1 \ {\rm keV}$, and the magnetic field scales
as $B\sim B_p(R/R_{NS})^{-3}$, where $B_p \sim 10 B_q$ is the
polar field and $R_{NS}$ is the star radius.

We assume that the magnetosphere is force free. Charges, flowing
along ${\mathbf B}$, are accelerated by the parallel electric
field which is provided by the decay of the twist, $\partial
(B^2/8\pi)/\partial t=-E_\parallel j$. A thermal photon scatters
where its energy in the particle frame,
$\gamma(1-\beta\cos\theta)\epsilon$, equals the local cyclotron
energy, $\epsilon_B = m_ec^2(B/B_Q)$ (here $\beta=v/c\sim 1$ and
$\theta$ is the angle between the incident photon and the magnetic
field), i.e. when the particle Lorentz factor is
$\gamma=\gamma_{res}\sim (m_ec^2/\epsilon)(B/B_Q)$. Because the
potential drop at the endpoints of a flux tube (of length $L$) is
enormous, $e\Phi_0 > 10^6 m_ec^2$ (see below), the electron
Lorentz factor, which is $\sim 1$ at the star surface,
monotonically increases up to a value which is much higher than
$\gamma_{res,max}\sim 500 (1\, {\rm keV}/\epsilon)(B_p/B_Q)\approx
5000$. This ensures that the resonant condition ($\gamma_{res} =
\gamma$) is met after the electron travelled a typical distance
$\lambda_{acc,res}=\gamma\vert dx/d(\gamma\beta)\vert\approx
\gamma_{res}L/(e\Phi_0/m_ec^2)\ll L$. On the other hand, this is
not a sufficient condition for RCS to occur: even if they can be
accelerated up to $\gamma_{res}$, electrons are able to scatter
only if $\lambda < \lambda_{acc,res}$, where $\lambda \sim
(\epsilon/m_ec^2)/(n_{ph}\sigma_{eff})\approx 10^{-3}(T/1\, {\rm
keV})^{-2}(R/R_{NS})^{2} \ {\rm cm}$ is the particle mean free
path for resonant scattering (here $n_{ph}$ is the density of
thermal photons and $\sigma_{eff}=3\pi\sigma_T/4\alpha_F$; e.g.
Dermer 1990). This translates into:
\begin{equation} \frac{e\Phi_0}{m_ec^2}<  5\times
10^{11}\left(\frac{B}{B_Q}\right)\left
(\frac{R}{R_{NS}}\right)^{-2}\left(\frac{T}
{1\,{\rm keV}}\right)^2 \left(\frac{L}{10^6\, {\rm cm}}\right)\, .
\label{eq1}
\end{equation}
We approximate the left hand quantity with the potential drop
which produces the conduction current in a 1D, relativistic double
layer (where $j\propto \Phi_0^2$; Carlqvist 1982):
\begin{equation}
\frac{e\Phi_0}{m_ec^2}\approx 2\times
10^{9}\sqrt\frac{m_+}{m_e}\left(\frac{B}
{B_Q}\right)^{1/2}\left(\frac{R}{R_{NS}}\right)^{-1/2}
\left(\frac{L}{10^6\, {\rm cm}}\right)\, , \label{phidl}
\end{equation}
where $m_+$ is the mass of the positive charge.
If the magnetosphere is populated by $e^-/e^+$ pairs then
$m_+=m_e$, and for a polar magnetic field $\sim 10B_Q$ the
inequality given by Eq.~(\ref{eq1}) is satisfied up to $B/B_Q
> 0.05$, or $R/R_{NS} < 6$.

The upscattered photon produced in the region $B/B_Q > 0.05$ has
an energy
$\epsilon'\sim\gamma_{res}^2\epsilon/(1+\gamma_{res}\epsilon/m_ec^2)$
and may convert into a $e^\pm$ pair in the strong magnetic field
provided that the threshold condition $\epsilon' >
2m_ec^2/\sin\theta'$ is met, where $\theta'$ is the angle of the
scattered photon with the magnetic field. For $\epsilon\sim 1\
{\rm keV}$, pair creation requires $\gamma_{res}
> 30$ (i.e. $B > 0.05B_Q$). Since electrons are relativistic, the scattered photon
initially moves parallel to $\mathbf B$ and a pair can be produced
only after a large enough pitch angle has built up. Moreover, pair
production is efficient only if its characteristic length-scale is
less than the dimension of the circuit, i.e. $\alpha_{\pm} >
1/R_{NS}$ where $\alpha_{\pm}$ is the absorption coefficient for
ordinary and extraordinary photons (see Baring  2007).
Accordingly, the limiting value of magnetic field above which the
process is effective decreases with increasing photon energy, and
it is $B\sim 0.05B_Q$ for pairs created near threshold.

In summary, in the region $B > 0.05 B_Q$ a quasi-equilibrium
configuration is reached with a pair multiplicity $\sim
L/\lambda_{acc,res}$ of a few. Screening of the electric field
limits the potential drop to
$e\Phi_0/m_ec^2\approx\gamma_{res}\sim 500(B/B_Q)$ and the maximum
$e^\pm$ Lorentz factor is $\gamma_{res}$. Charges undergo only few
scatterings with thermal photons, but they loose most of their
kinetic energy in each collision. A steady situation is maintained
against such Compton losses because $e^-/e^+$ are re-accelerated
by the electric field before they can scatter again. RCS may occur
also at larger radii provided that the charges Lorentz factor is
limited to $\sim \gamma_{res}$, which decreases as $B\sim 1/R^3$.
The pairs energy depends on the electrostatic potential (in turn
fixed by conduction current), and also on the efficiency of the
radiative (Compton) drag. If pairs with $\gamma\sim \gamma_{res}$
are injected in the external region ($B/B_Q < 0.05$), the circuit
is likely to behave much differently from a double layer, allowing
the current to be conducted with only a small potential drop (see
also Beloborodov, this volume).

\section{IR/Optical emission}
\label{sec:3}

In order to estimate the amount of curvature emission from the
inner magnetosphere, as seen by a distant observer, we use a
simple geometrical model. We approximate the twisted field with a
dipole, ${\mathbf
B}=B_p(R/R_{NS})^{-3}(\cos\theta,\sin\theta/2,0)$, while the
current is still taken to be $j=(c/4\pi)|\nabla\times {\mathbf
B}|_{twist}\approx B(R/R_{NS})^{-1}$. The particle density is then
$n_\pm= B(R,\theta)\left(R/R_{NS}\right)^{-1}/\left (4 \pi e
\right )$, for a pair velocity $\simeq c$.  With reference to a
generic flux tube labelled by $\theta_0$
($0\leq\theta_0\leq\pi/2)$, curvature photons will reach the
observer if they are emitted in the volume $\Delta V= X
R_c\sin\theta_{obs} \left\vert\
dZ/d\theta_0\right\vert\Delta\theta_0/   \gamma^2$, around the
point $P$ of cartesian co-ordinates ($X,\, Z$) where $\mathbf B$
is parallel to the LOS (Fig.\ref{volumi}). Here $\theta_{obs}$ is
the viewing angle,i.e. the angle between the line-of-sight (LOS)
and the magnetic axis. In general there are two emitting regions
along each flux tube one above the other below the magnetic
equator, because of electrons and positrons, and for any
$\theta_0$ there are two azimuthally symmetric flux tubes.

The (polarization averaged) CR spectral power per particle can be
expressed as $ p(\epsilon)=\left (\dot E/\epsilon_c \right ) \left
[ (\epsilon/\epsilon_c) \int_{\epsilon/\epsilon_c}^\infty
K_{5/3}(x)dx  \right ] / \left [\int_0^\infty xdx\int_x^\infty
K_{5/3}(x')dx'\right ]$, where $K_{5/3}(x)$ is the modified Bessel
function, $\epsilon_c=(3/2)\hbar c\gamma^3/R_c$, $\dot
E=(2/3)e^2c\gamma^4/R_c^2$, and $R_c$ is the curvature radius of
the field line. The monochromatic luminosity emitted by the volume
$\Delta V$ is $\Delta L=p(\epsilon)n_\pm \Delta V$, and, since
radiation is collimated in the solid angle
$\Delta\Omega=\pi/\gamma^2$, the observer at infinity measures a
luminosity $4\pi \Delta L/\Delta\Omega$. The total received power
is then obtained summing the individual contributions over all the
flux tubes, assuming the pair Lorentz factor is locked at
$\gamma_{res}$.

% For figures use
%
\begin{figure}[b]
%\sidecaption
% Use the relevant command for your figure-insertion program
% to insert the figure file.
% For example, with the graphicx style use
\includegraphics[scale=.35]{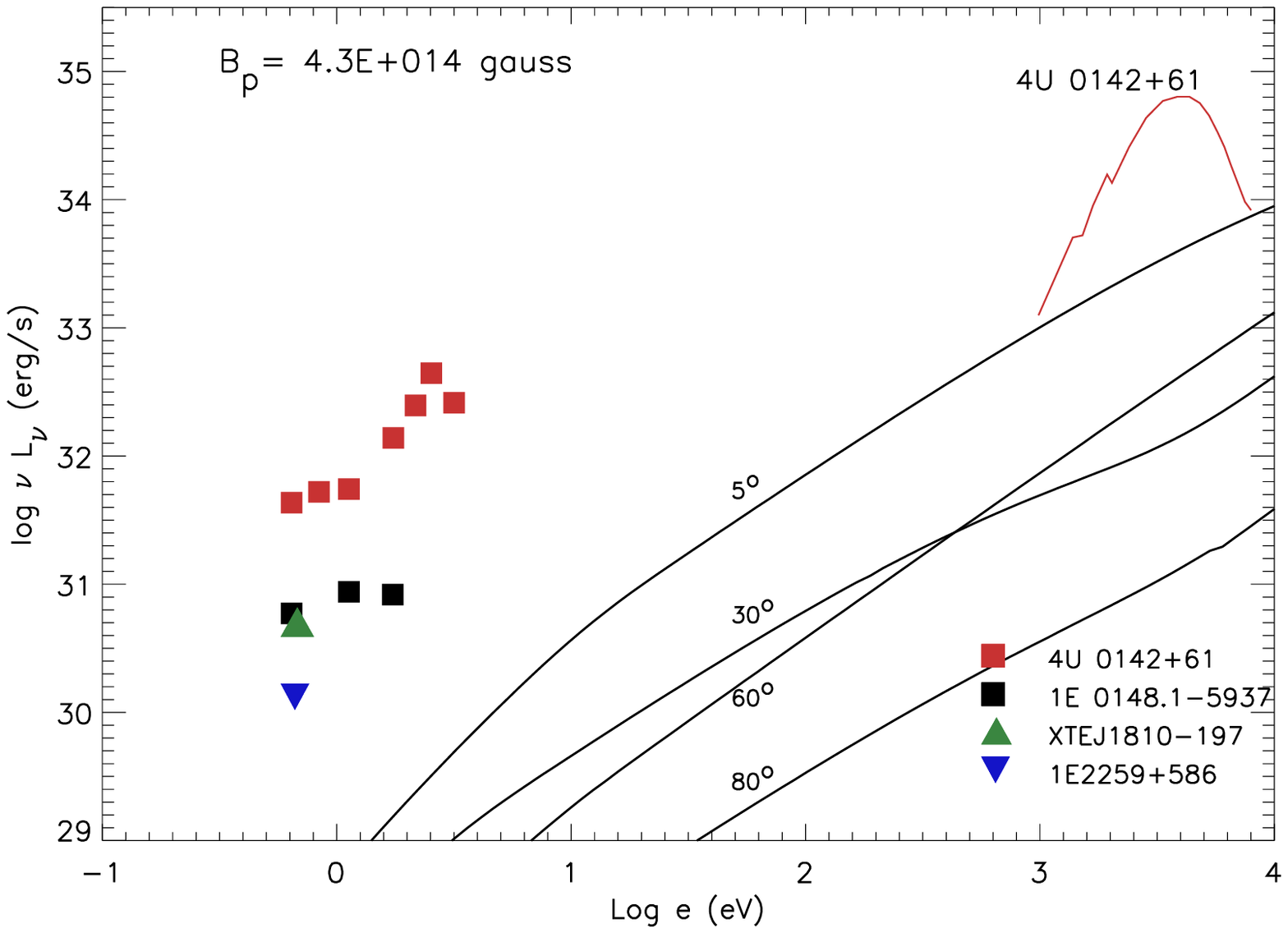}
\includegraphics[scale=.35]{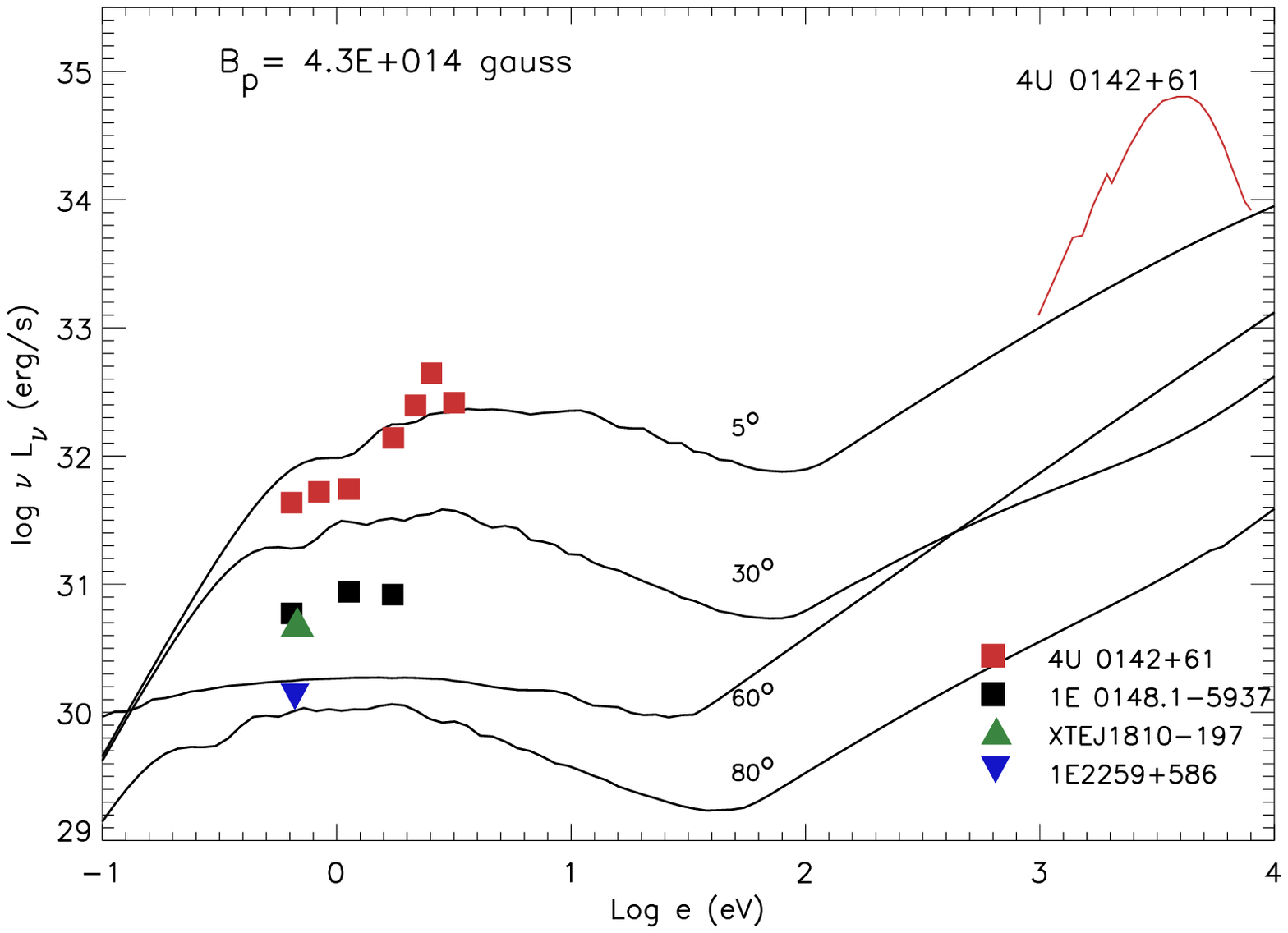}
\caption{ Model spectra for different values of $\theta_{obs}$ and
$B_p=4.3\times 10^{14}$~G. The {\it XMM-Newton} X-ray spectrum of
\rxs\ is from Rea et al. (2008; red solid line). The
AXPs IR/optical data are from Duncan \& van Kerkwijk 2005 (\uu\
and \ee) and Mignani et al. 2007 (\xte\ and \ea). The adopted
distances for de-reddening are 5~kpc (\uu, \rxs), 3~kpc (\ee,
\ea), 4~kpc (\xte), 8.5~kpc (\kes). {\em Left:} Spectra computed
for incoherent curvature emission. {\em Right:} Same as in the
left panel, but accounting for particle bunching.}
\label{spectra1}
\end{figure}

Fig. \ref{spectra1} (left panel) shows the computed CR spectrum
for $B_p=4.3\times 10^{14}$ and different viewing angles
$\theta_{obs}$, together with the optical/IR luminosities observed
for a sample of AXPs. It should be noted that all these sources
typically show variability in all energy bands. The data shown in
Fig.~\ref{spectra1} were not selected according to any particular
criterion and are just meant to be representative of the IR
emission of magnetar candidates. As it appears from
Fig.~\ref{spectra1}, model spectra severely underpredicts
optical/IR data. Basically, this is due to the fact that
the particle density is too low for incoherent CR emission to
produce the required power. The point has been already noted, on
the basis of a qualitative analysis, by BT07 who concluded that
particle bunching would be likely present, providing the required
amplification of CR emission.

Many models have been suggested to explain the origin of the
interactions which push particles together and can lead to the
formation of bunches of charged particles localized in
phase-space, but the most promising explanation seems to be
connected with plasma instabilities, like the two-stream
(electron-positron/electron-ion) instability. If $N$ particles are
contained in a bunch of spatial scale $l_B$, they radiate like a
single particle of charge $Q=Ne$. This collective emission
produces an amplification of the radiated power by a factor $N$
with respect to the standard, non-coherent emissivity by $N$
individual charges. The details of the emission depend on the
shape of the bunch. In the simplest, one-dimensional case the
power per unit frequency emitted by a bunch is $p_B(\nu)\sim
N^2p(\nu)\sin^2(\pi\nu/\nu_l)/(\pi\nu/\nu_l)^2=N^2{\cal P}(\nu)$
(Saggion 1975). The main source of uncertainties lies in the size
of the bunch, and here we assume that $l_B$ is associated to the
local plasma frequency, $l_B\sim c/\nu_{pe}= \pi m_ec^2/(\gamma
e^2n_e)$ (e.g. Lesch 1998). Since a single bunch emits a spectral
power $(n_el_B^3)^2{\cal P}(\nu)$ and there are $N_B\sim\Delta
V/l_B^3$ bunches in each emitting volume, the total spectral power
radiated within $\Delta V$ is $\Delta L_B=N_B(n_el_B^3)^2{\cal
P}(\nu)=(n_el_B^3)\Delta L$. We stress that it has to be
$N=n_el_B^3>1$ for the process to be effective, meaning that only
emission at frequencies $\nu < \nu_l<\nu_{co}=n_e^{1/3}c$ is
efficiently amplified. Where $N<1$ emission is from individual
particles. Finally, because of the strong absorption below the
plasma frequency, a low-energy cut-off is present around
$\nu_{pe}$.

We have recalculated the emergent spectrum including particle
bunching (Fig.~\ref{spectra1}, right panel), finding that coherent
curvature radiation can indeed produce sufficient IR/optical
emission to account for the observed one. Given the limitations of
our model, we are not in a position to attempt any fit of the
data. What we aim to, instead, is verifying that magnetospheric CR
emission is able of reproduce the gross characteristics of the
observed low-energy emission, its energetics in particular. The
predicted spectrum vary with the viewing angle, and the luminosity
is higher when the star is viewed nearly pole-on. The curvature
spectrum is far below the observed one in the soft X-rays,
indicating that a different mechanism is necessary to account for
the X-ray emission below $\sim 10$~keV. In the twisted
magnetosphere scenario this is RCS onto mildly relativistic pairs
populating the external magnetosphere.

\section{Discussion and Conclusion}
\label{sec:4}

Despite the magnetar scenario is now regarded as the most likely
interpretation for the observed properties of SGRs and AXPs, and
the many investigations aimed at producing detailed models for the
spectral and timing properties of magnetars, a number of key
issues are still unresolved. In particular, the origin of the
low-energy emission (in the IR/optical band) of a magnetar is
still under debate. In this paper we gave a quantitative estimate
of the IR/optical emission, under the assumption that it arises
from curvature radiation from pairs in the inner magnetosphere. A
comparison with observations of AXPs shows that the predictions of
the model are in general agreement with observations, provided
that curvature radiation is not coherent and a bunching mechanism
is at work at long wavelengths.

Our model is based on a number of simplifying assumptions. For
instance, the computation is based on a globally twisted dipolar
magnetosphere while in some magnetars, the transient AXPs
in particular, the twist might affect only a limited bundle of
field lines close to a magnetic pole (Beloborodov 2009) . If the
inner part of the magnetosphere is (or becomes) untwisted, no
currents are present that can produce the IR/optical 
curvature radiation. Also, we used a simple double layer model (eq.~1) for 
the linear accelerator, which may not be valid in presence of intense pair 
creation (BT07). Furthermore, and more important, the details of the
charges motion in the external region of a twisted magnetosphere,
which are essential ingredients in providing a model for the high
energy emission in the soft and hard X-ray band ($\sim 0.5$--200
keV), are still unclear. In order to reach firmer conclusions
about the entire multi-wavelength spectrum a more detailed study
of the magnetosphere is required, and will be matter of future
work (see also Beloborodov, this volume).

\begin{acknowledgement}
LN and RT are partially supported by the INAF/ASI grant
AAE-I/088/06/0. The authors thank the organizers of the Sant Cugat
forum for the hospitality and support.
\end{acknowledgement}

\input{ref}

\end{document}

%% file: ref.tex
%%%%%%%%%%%%%%%%%%%%%%%% referenc.tex %%%%%%%%%%%%%%%%%%%%%%%%%%%%%%
% sample references
% %
% Use this file as a template for your own input.
%
%%%%%%%%%%%%%%%%%%%%%%%% Springer-Verlag %%%%%%%%%%%%%%%%%%%%%%%%%%
%
% BibTeX users please use
% \bibliographystyle{}
% \bibliography{}
%